%% file: main.tex
\documentclass{article}\usepackage[left=3cm, right=3cm, top=2cm, bottom = 2cm]{geometry}

\usepackage{graphicx}
\usepackage{url}
\usepackage{array}
\makeatletter
\newcommand{\@chapapp}{\relax}%
\makeatother
\usepackage[title,header]{appendix}

\usepackage[xindy]{glossaries}
\loadglsentries{acronyms}
\usepackage{hyperref}
\usepackage{wrapfig}
\usepackage{paralist}
\usepackage{microtype}
\usepackage{lscape}
\usepackage{textcomp}
\usepackage{siunitx}
\usepackage{tabularx}

\usepackage{fretish} 

\begin{document}

\title{FRETting about Requirements: Formalised Requirements for an Aircraft Engine Controller\thanks{
The authors thank Georgios Giantamidis, Stylianos Basagiannis, and Vassilios A. Tsachouridis (UTRC, Ireland) for their help in requirements elicitation; and Anastasia Mavridou (NASA Ames Research Center, USA) for her help with FRET. 
This research was funded by the European Union’s Horizon 2020 research and innovation programme under the VALU3S project (grant No 876852), and by Enterprise Ireland (grant No IR20200054). The funders had no role in study design, data collection and analysis, decision to publish, or preparation of the manuscript.}}



\author{Marie Farrell \and Matt Luckcuck \and Ois\'{i}n Sheridan \and Rosemary Monahan}

\date{Department of Computer Science, Maynooth University, Ireland\\
\texttt{valu3s@mu.ie}}

\maketitle



\begin{abstract}
\textbf{[Context \& motivation]} Eliciting requirements that are detailed and logical enough to be amenable to formal verification is a difficult task. Multiple tools exist for requirements elicitation and some of these also support formalisation of requirements in a way that is useful for formal methods.  
\textbf{[Question/problem]} This paper reports on our experience of using the \gls{fret} alongside our industrial partner. The use case that we investigate is an aircraft engine controller. In this context, we evaluate the use of \gls{fret} to bridge the communication gap between formal methods experts and aerospace industry specialists. 
\textbf{[Principal ideas/results]} We describe our journey from ambiguous, natural-language requirements to concise, formalised \gls{fret} requirements. We include our analysis of the formalised requirements from the perspective of patterns, translation into other formal methods and the relationship between parent-child requirements in this set. We also provide insight into lessons learned throughout this process and identify future improvements to \gls{fret}. 
\textbf{[Contribution]}  Previous experience reports have been published by the \gls{fret} team, but this is the first such report of an industrial use case that was written by researchers that have not been involved \gls{fret}'s development.\end{abstract}

\glsresetall

\section{Introduction}
\label{sec:intro}

Formal verification uses mathematically-based techniques to guarantee that a system obeys certain properties, which is particularly useful when developing safety-critical systems like those used in the aerospace domain. Developing a correct set of requirements necessitates discussion with people who have expertise in the system under development, who may not have skills in formal methods. In which case, it can be beneficial to the requirements elicitation process to write the requirements in an intermediate language. Tools like NASA's \gls{fret} provide a gateway for developing formal requirements with developers who are not familiar with formal languages~\cite{giannakopoulou_formal_2020}. 


In this paper, we examine how \gls{fret} can be used in an industrial case study of an aircraft engine controller that has been supplied by our industrial partner, United Technologies Research Center (Ireland). \gls{fret} has previously been used to formalise the requirements for the 10 Lockheed Martin Cyber-Physical Challenges~\cite{mavridou2020ten}. However, to the best of our knowledge this paper provides the first experience report on \gls{fret}'s use on an industrial case study, by a team not involved in \gls{fret}'s development.

Our approach provides external and internal traceability. Using a tool like \gls{fret} to develop the requirements provides a link between natural-language requirements and formally verified artefacts. \gls{fret} also enables the user to describe a link between requirements at different levels of abstraction, which means that this traceability is maintained within the developing set of requirements. We also use \gls{fret} to collect information about the rationale behind a requirement, further improving the traceability; either back to a natural-language requirement, or forward to a more concrete requirement. These traceability features encourage better explainability of a requirement's source, and the intermediate language improves the explainability of the requirements themselves.


The rest of the paper is laid out as follows.
\S\ref{sec:bg} outlines the relevant background material pertaining to \gls{fret} and the aircraft engine controller use case. Then, we describe our requirements elicitation process and present detailed requirements in \S\ref{sec: process}. These requirements are analysed in \S\ref{sec: analysis}. We discuss the lessons that were learned through this work in \S\ref{sec: lessons} and \S\ref{sec:conclude} concludes. We also provide a detailed appendix showing the full set of requirements, test cases, Simulink model and \fretish{} requirements.

\section{Background}
\label{sec:bg}
This section provides an overview of \gls{fret} and the aircraft engine controller use case for which we were developing requirements. 

\paragraph{FRET:}
\label{sec:fretIntro}

is an open-source tool that enables developers to write and formalise system requirements~\cite{giannakopoulou_formal_2020}. \gls{fret} accepts requirements written in a structured natural-language called \fretish{}, in which requirements take the form:\\ 
\centerline{\fretishComponents{}}
\noindent The \Condition{}, \Component{}, and \Response{} fields are mandatory; \Scope{} and \Timing{} are optional fields. This allows \Response{}s that are tied to a \Scope{}, are triggered by \Condition{}s, relate to a system \Component{}, and may have \Timing{} constraints.

The underlying semantics of a \fretish{} requirement is determined by the \Scope{}, \Condition{}, \Timing{}, and \Response{} fields. There is a template for each possible combination of a requirement's fields, currently \gls{fret} provides 160 such templates \cite{giannakopoulou2021automated}. The selected template is used to generate formalisations  of the associated requirement in both past- and future-time metric \gls{ltl}. 
\gls{fret} displays a diagramatic semantics for each requirement, which shows: the time interval where it should hold, and its triggering and stopping conditions (if they exist). Both versions of the requirements are helpful for sanity-checking what has been written in \fretish{}.

The user must give each \fretish{} requirement an ID, which can be used to create a many-to-many, hierarchical link between requirements: a \textit{parent} requirement may have many \textit{child} requirements, and one child may have many parents. While this link facilitates traceability, \gls{fret} does not define this relationship (formally or otherwise). For example, a child requirement does not inherit definitions from its parent. We discuss possible improvements to this link in \S\ref{sec:analysisParentChild}. 
\gls{fret} also allows the user to enter `Rationale' and `Comments' for a requirement, which further supports traceability and encourages explainability of requirements.

\gls{fret} can automatically translate requirements into contracts for a Simulink diagram, written in CoCoSpec, which are checked during Simulink simulations by the CoCoSim tool, using the Kind2 model checker~\cite{bourbouh2020cocosim}. \gls{fret} can also generate runtime monitors for the Copilot framework~\cite{dutle2020requirements}.

\paragraph{Aircraft Engine Controller:}
\label{sec:controllerIntro}

Our use case is a software controller for a high-bypass civilian aircraft turbofan engine, provided by our industrial partner on the VALU3S~\cite{barbosa2020valu3s} project, based on existing controller designs~\cite{postlethwaite1995digital,samar_design_2010}. It is an example of a Full Authority Digital Engine Control (FADEC) system, which monitors and controls everything about the engine,
using input from a variety of sensors. The engine itself contains two compressors (high-pressure and low-pressure) turning a central spool, which drives the engine.

As described in our prior work~\cite{luckcuck2021verifiable}, the controller's high-level objectives are to manage the engine thrust, regulate the compressor pressure and speeds, and limit engine parameters to safe values. It should continue to operate, keeping settling time, overshoot, and steady state errors within acceptable limits, while respecting the engine's operating limits (e.g. the spool's speed limit), in the presence of:
\begin{compactitem}
\item sensor faults (a sensor value deviating too far from from its nominal value, or being unavailable),
\item perturbation of system parameters (a system parameter deviating  too far from from its nominal value), and
\item other low-probability hazards (e.g. abrupt changes in outside air pressure).
\end{compactitem}
\noindent The controller is also required to detect engine surge or stall and change mode to prevent these hazardous situations. 
 

Our industrial partner has supplied us with 14 English-language requirements (Table~\ref{table: nlreqs}) and 20 abstract test cases, which provide more detail about the controller's required behaviour. 
The naming convention for requirements is:\\ \centerline{$<$\textit{use case id}$>$\_R\_$<$\textit{parent requirement id}$>$.$<$\textit{child requirement id}$>$}\\
For example, because this is Use Case 5 in the VALU3S project\footnote{The VALU3S project: \url{https://valu3s.eu/}}, requirement one is named UC5\_R\_1. Note, we use a similar naming convention for test cases. Table~\ref{table:tcr1} shows the abstract test cases for UC5\_R\_1. Our industrial partner also designed the controller in Simulink\footnote{Simulink: \texttt{\href{https://mathworks.com/products/simulink.html}{https://mathworks.com/products/simulink.html}}}, shown in the extended version of this paper\footnote{This Paper Extended Version: \url{https://arxiv.org/abs/2112.04251}}.

For our Use Case, we collaborated with scientists in the
System Analysis and Assurance division 
of an aerospace systems company. 
The hour-long requirements elicitation meetings were held monthly, over a period of 10 months, with additional meetings as needed.
In these meetings, our collaborators reviewed the \fretish{} versions of their natural-language requirements, validating our formalisation and clarifying ambiguities for us. 
Since our collaborators were already familiar with other formal tools we were able to introduce them to \gls{fret} quite quickly. However, we produced a training video for other members of the project consortium\footnote{``Formalising Verifiable Requirements'' Presentation: \url{https://www.youtube.com/watch?v=FQGKbYCbxPY&list=PLGtGM9euw6A66ceQbywXGjVoTKEhP-Of7&index=9}}.

\section{Our Requirements Elicitation Process Using FRET}
\label{sec: process}

In this section we describe our requirements elicitation process. We begin by outlining how this fits into our larger approach to verification for the aircraft engine controller use case. We then describe our journey from natural-language requirements to formalised \fretish{} requirements.

\subsection{Requirements-Driven Methodology}
\label{subsec:methodology}

\begin{figure}[t]
\centering
\includegraphics[width=\textwidth]{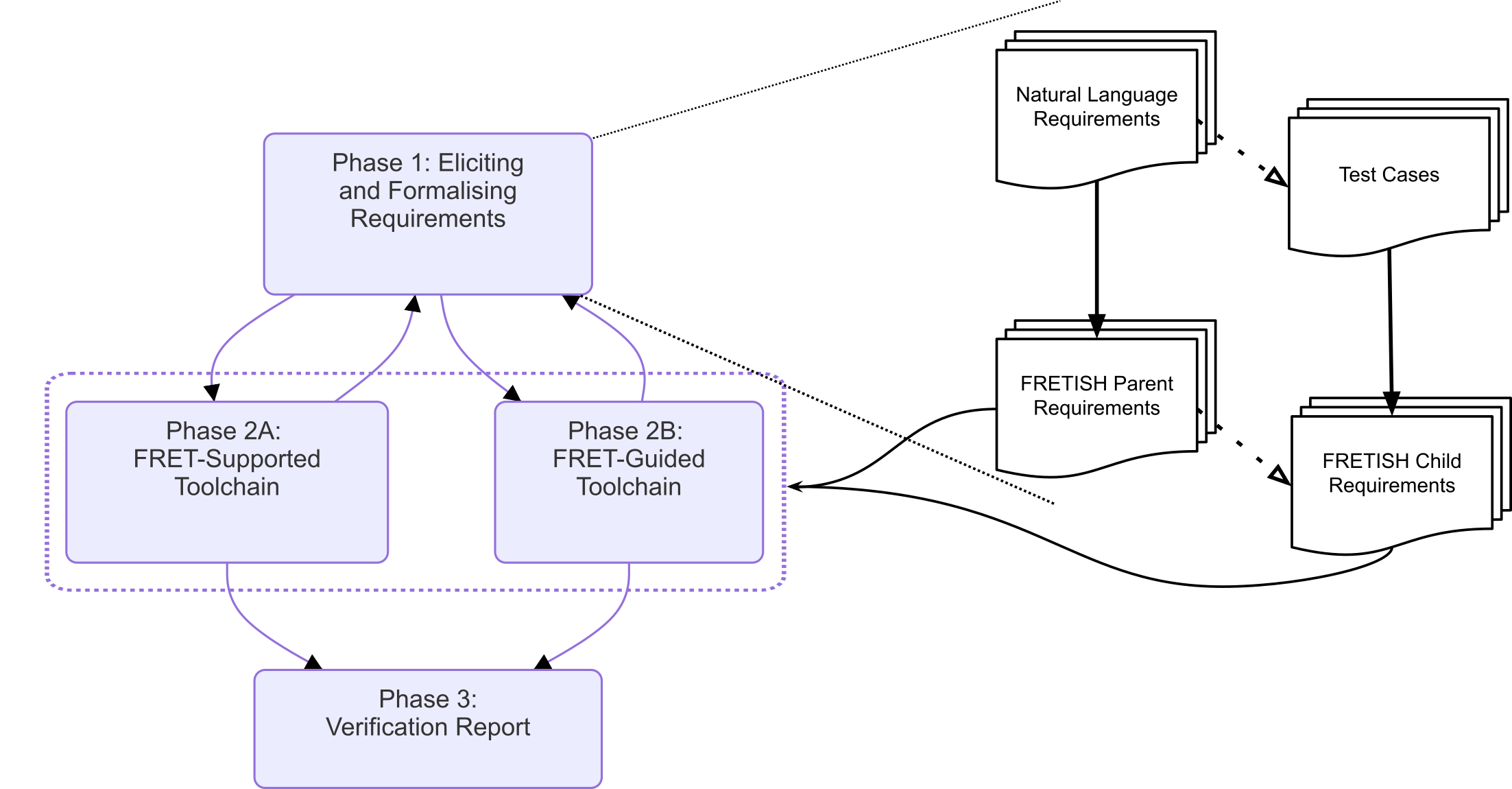}
\caption{Flowchart of our three-phase requirements-driven verification methodology~\cite{luckcuck2021verifiable} (left) with an exploded view of Phase 1's artefacts (right). The solid lines and arrowheads show direct information flow between artefacts (and back into Phase 2A and 2B), the dashed lines and open arrowheads show an artefact being implemented by another.}
\label{fig:methodology}
\end{figure}

As part of the three-phase verification methodology outlined in our prior work~\cite{luckcuck2021verifiable}, we used \gls{fret} to elicit and formalise requirements for the aircraft engine controller. Focussing on Phase 1, this paper includes the full set of \fretish{} requirements and presents lessons learnt, which were not discussed in \cite{luckcuck2021verifiable}. 
 Fig.~\ref{fig:methodology} shows a high-level flowchart of our methodology, with an exploded view of the relationship between the artefacts involved in Phase 1.  The methodology takes the natural-language requirements (Table~\ref{table: nlreqs}), test cases, and Simulink diagram of the engine controller as input, and enables the formal verification of the system's design against the requirements. 

Phase 1 of our methodology involves formalising natural-language requirements using \gls{fret}, eliciting further detail as we progress. Phase 2 consists of two, potentially parallel, phases. Phase 2A uses \gls{fret}'s built-in support for generating CoCoSpec contracts that can be incorporated into a Simulink diagram for verification with the Kind2 model-checker. In Phase 2B, the formalised requirements drive a (manual) translation into other formal methods, as chosen by the verifier. These tools typically require the construction of a formal model of the system, which is verified against the translated requirements. This step requires translation of the \fretish{} requirements into the formalism of the chosen verification tool.  
Finally, Phase 3 produces a report collecting the verification results and other useful artefacts, such as formal models, Simulink diagrams, various versions of the requirements, counter-examples, proofs, etc. This supports tracing the requirements through the system's development lifecycle.

The following subsections describe the requirements elicitation process (Phase 1) in more detail. 
Figure~\ref{fig:methodology}'s exploded view, shows how natural-language requirements are translated into \fretish{} parent requirements (solid arrow), and the test cases are translated into  child requirements. Since we view the test cases as implementations of the natural-language requirements (dashed arrow), the child requirements are similarly viewed as implementations of their corresponding parent requirements.
The left-hand side of Fig.~\ref{fig:methodology} shows how the work in this paper fits within our development and verification methodology; the solid arrows from the \fretish{} parent and child requirements, to Phases 2A and 2B, show how the output of this work is consumed by the next phase.

\input{nlreqtable}

\begin{table}[t]
\begin{tabular}{|m{1.5cm}|m{2.2cm}|p{10.5cm}|}
\hline
\textbf{Test Case ID} & \textbf{Requirement ID} &\textbf{Description} \\ \hline
UC5\_TC\_1 & UC5\_R\_1 & \textbf{Preconditions}: Aircraft is in operating mode M and sensor S value deviates at most +/- R \% from nominal value \newline
\textbf{Input conditions/steps}: Observed aircraft thrust is at value V1 and pilot input changes from A1 to A2 \newline
\textbf{Expected results}: Observed aircraft thrust changes and settles to value V2, respecting control objectives (settling time, 
overshoot, steady state error) \\ \hline
UC5\_TC\_2 & UC5\_R\_1 & \textbf{Preconditions}: Aircraft is in operating mode M and sensor S value is not available (sensor is out of order) \newline
\textbf{Input conditions/steps}: Observed aircraft thrust is at value V1 and pilot input changes from A1 to A2 \newline
\textbf{Expected results}: Observed aircraft thrust changes and settles to value V2, respecting control objectives (settling time, 
overshoot, steady state error)\\\hline
\end{tabular}
\caption{Abstract test cases corresponding to requirement UC5\_R\_1. Each specifies the preconditions for the test case, the input conditions/steps and the expected results.}
\label{table:tcr1}
\end{table}

\subsection{Speaking FRETISH: Parent Requirements}
\label{sec: fretreqs}

\input{frethlreqs}

The inputs to our requirements elicitation process were the Simulink diagram, 14 natural-language requirements (Table \ref{table: nlreqs}), and 20 abstract test cases that were supplied by our industrial partner. We elicited further information about the requirements through regular team discussions with our industrial partner.

We started by translating the natural-language requirements into \fretish, producing the set of 14 \fretish{} requirements in Table \ref{table: frethlreqs}. 
The correspondence between the \fretish{} requirements and their natural-language counterparts is clear. For example, requirement UC5\_R\_1 states that: \smallskip
\begin{compactitem}
\item[]\textit{Under sensor faults, while tracking pilot commands, control objectives shall be satisfied (e.g., settling time, 
overshoot, and steady state error will be within predefined, acceptable limits).}
\end{compactitem}
\smallskip

\noindent This became the corresponding \fretish{} requirement:\smallskip
\begin{compactitem}
\item[]\condition{if ((sensorfaults) \& (trackingPilotCommands))} \component{Controller} shall\\ \response{(controlObjectives)}
\end{compactitem}
\smallskip

\noindent Producing this initial set of requirements enabled us to identify the ambiguous parts of the natural-language requirements. For example, the phrase ``\textit{sensor faults}'' simply becomes a boolean in our \fretish{} requirements, highlighting that we need to elicit more details. We captured these additional details as child requirements, as described in \S\ref{sec:children}

\subsection{Adding Detail: Child Requirements}
\label{sec:children}

Once the \fretish{} parent requirements (Table~\ref{table: frethlreqs}) were complete, we added more detail to make the requirements set more concrete. We paid particular attention to ambiguous phrases translated from the natural-language requirements. These extra details were drawn from the abstract test cases and from detailed discussions with our industrial collaborators, who clarified specific ambiguities. 

We captured the extra details in 28 child requirements. As mentioned in \S\ref{sec:fretIntro}, a child requirement does not inherit definitions from its parent(s). However, we use this hierarchical link to group the detail in the child requirements under a common parent, which enables the detailed child requirements to be traced back to the more abstract parent requirements.

For example, UC5\_R\_1 was distilled into three requirements (UC5\_R\_1.1, UC5\_R\_1.2 and UC5\_R\_1.3), shown in Table \ref{table: r1children}. These three child requirements each have the same \Condition{} and \Component{}, but differ in their \Response{}s.
Each child requirement specifies one of the ``\textit{control objectives}'' (settling time, overshoot and steady state error) mentioned in the natural-language version of UC5\_R\_1. During elicitation discussions, it was revealed that these were actually the \textit{only} control objectives that were of concern for this use case. Here, using \gls{fret} encouraged us to question exactly what the phrase ``\textit{control objectives}'' meant.

Each of these requirements includes the condition \condition{when (diff(r(i),y(i)) $>$ E)} and the timing constraint \timing{until (diff(r(i),y(i)) $<$ e)}, which were initially overlooked in the natural-language requirements but revealed during elicitation discussions with our industrial partner.
The \Response{} must hold when the difference between the reference sensor value, $r(i)$, and the observed sensor value, $y(i)$, falls between specific bounds ($E$ and $e$). This important detail was  missing from the parent requirement but was uncovered during our requirements elicitation.

The ``\textbf{Preconditions}'' of test cases UC5\_TC\_1 and UC5\_TC\_2 (Table \ref{table:tcr1}) showed us that the phrase ``\textit{Under sensor faults}'' meant a period where a sensor value deviates by \textpm R\% from its nominal value or returns a \texttt{null} value. To represent this, the child requirements use the function  \condition{sensorValue(S)} where \condition{S} is a parameter representing each of the 4 sensors for the engine controller. These requirements are thus applied to all of the sensors in the model.

In UC5\_TC\_1 and UC5\_TC\_2, the ``\textbf{Input conditions/steps}'' refer to the aircraft thrust and a change in the pilot's input. We encoded this as the condition and response pair \condition{(pilotInput $=>$ setThrust = V2) \& (observedThrust = V1)} and \response{(observedThrust = V2)}, where \condition{V1} and \condition{V2} are thrust variables and \condition{=>} is logical implication. During elicitation discussions we found that this pair corresponds to the \Condition{}, \condition{trackingPilotCommands}. This was a particularly important clarification because \condition{trackingPilotCommands} models the phrase ``\textit{while tracking pilot commands}'', which the natural-language requirements use extensively. This underlines that it is possible for an ambiguous statement to have a very precise meaning that was lost while drafting the requirements.

The thrust variables \condition{V1} and \condition{V2} in our \fretish{} requirements correspond to variables \texttt{V1}, \texttt{V2}, \texttt{A1}, and \texttt{A2} in the test cases. During elicitation discussions, we found that \texttt{V1} and \texttt{V2} alone were sufficient to capture the requirement. \texttt{V1} and \texttt{A1} are used interchangeably as the initial thrust value, which we label \condition{V1}. Similarly, \texttt{V2} and \texttt{A2} refer to the updated thrust value, which we label \condition{V2} for consistency.
 This is another ambiguity that our translation from natural-language to \fretish{} helped to clarify. 

Our industrial partner checked the child requirements to ensure that there were no errors or omissions. The intuitive meaning of \fretish{} constructs simplified this check, and features like the requirements' diagramatic semantics provided quick feedback when we edited the requirements during elicitation discussions. The act of formalising the requirements helped us to identify ambiguities in the requirements, prompting elicitation of further detail from our industrial partner.


\input{r1childreqs}

\section{An Analysis of Elicited Requirements}
\label{sec: analysis}

This section provides an analysis of the \fretish{} requirements that we produced for the aircraft engine controller use case. We note that the requirements only refer to one component, the \component{Controller}, but this could be decomposed to refer to specific blocks in the use case Simulink design.

\subsection{Requirement Templates}
\label{sec: patterns}

Each of the 14 \fretish{} parent requirements (Table \ref{table: frethlreqs}) uses the same pattern: \Condition{} \Component{} shall \Response{}. As described in \S\ref{sec:fretIntro}, \gls{fret} maps each requirement into a semantic template so that it can generate the associated LTL specification. Our parent requirements all correspond to the template [\textit{null, regular, eventually}], which specifies the \textit{scope-option}, \textit{condition-option} and \textit{timing-option}, respectively (if the \textit{timing-option} is omitted, then \textit{eventually} is the default). Specific details about templates in \gls{fret} are given in \cite{giannakopoulou2021automated}. 

We introduced \timing{until} clauses into all of the 28 child requirements, although with different timing constraints. The introduction of the \timing{until} clauses was identified through a combination of the information in the test cases and from extensive discussions with our industrial parter. However, the specific timing constraints required
in-depth discussion with our industrial partner to identify. Most of the child requirements correspond to the template [\textit{null, regular, until}]. However, some child requirements differed slightly as outlined below.

\input{r1314children}

UC5\_R\_13 and UC5\_R\_14 generated a lot of discussion, because they differ so much from the other requirements; here, the system changes between modes of operation, so we use the \Scope{} clause. This produced the child requirements shown in Table~\ref{table:fret1314children}. The \condition{when} and \timing{until} clauses differ from the other requirements, because here the mode change is triggered by comparing the set value of the low-pressure compressor's spool speed (\condition{setNL}) to the value produced by the sensor (\condition{observedNL}). It is necessary to differentiate between the cause of the difference, i.e. whether it was directly caused by \condition{pilotInput} or by external factors (\condition{!pilotInput}). In either case the system must change modes, but our industrial partner felt that it was important that the requirements distinguish the difference. 

\begin{figure}[t]
\centering
\includegraphics[width = 0.9\textwidth]{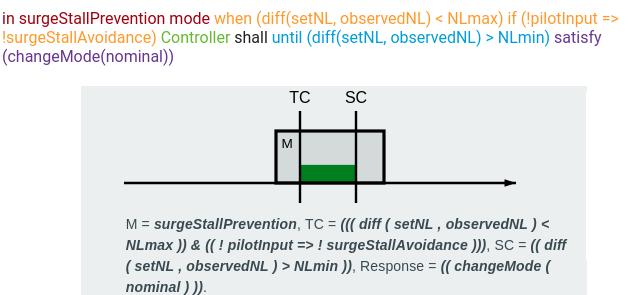}
\caption{\fretish{} and semantics diagram generated for UC5\_R\_14.2. Here, `M' indicates the mode, `TC' the triggering condition, and `SC' the stopping condition.}
\label{fig: fretr14.2}
\end{figure}

Fig.~\ref{fig: fretr14.2} contains the semantics diagram produced by \gls{fret} for UC5\_R\_14.2. The semantic template that was used is [\textit{in, regular, until}]. In a recent study using \gls{fret} to formalise the 10 Lockheed Martin Cyber Physical Challenge problems, the most commonly used semantic template was [\textit{null, null, always}] \cite{mavridou2020ten}. Of these 10 problems, the autopilot system is the closest to our case study, and it was the only requirement set to use the \textit{in} scope-option. The timing-option in their requirements was different to ours; but we use \timing{until}, which was introduced into \fretish{} after that study.


We used all of the fields available in \fretish{} in our use case, although a lot of our individual requirements used a subset of them. We only used \Scope{} in the four child requirements of UC5\_R\_13 and UC5\_R\_14. \gls{fret} provides many ways of specifying modes, but we only used \scope{} for this; there are many ways to specify a \Condition{}, but we only used \condition{when} and \condition{if}. There are also multiple ways to specify \Timing{}, but in this case study we only used \timing{until} clauses. 

Despite \timing{until} providing timing constraints, we did not use explicit times (e.g. ticks/timesteps/seconds) in our requirements. This is because the natural-language requirements (Table \ref{table: nlreqs}) do not mention timing, and our introduction of timing constraints came from elicitation discussions. However, time points are implicit in some of the child requirements, e.g. comparing \condition{r(i)} and \condition{y(i)} in the child requirements of UC5\_R\_1(Table~\ref{table: r1children}), or the \condition{T1} and \condition{T2} variables in UC5\_R\_11.1 (Table \ref{table: r11.1}). The \Timing{} clause was not intentionally avoided, but we felt that the implicit time constraints better suited the requirements and was closer to the description agreed with our industrial partner.

\input{r11_1}

\subsection{Parent-Child Relationship in our Use Case}
\label{sec:analysisParentChild}

As previously mentioned, \gls{fret} allows a requirement to be related to another as a `parent' or a `child', but this relationship is not well defined, formally or otherwise.
The parent-child relationship in \gls{fret} could be viewed as formal refinement~\cite{back_refinement_1998}: a concept supported by a variety of formal methods that enable formal specifications to be gradually made more concrete, while proving that they still obey a more abstract version of the specification. Similar approaches exist in the literature on refactoring goal-driven requirements~\cite{darimont1996formal,zave1997four}. 

If viewed through the lens of refinement, we would need to introduce \textit{abstraction invariants} to relate the abstract and concrete specifications. These invariants facilitate the proof that the concrete specification does not permit any behaviours that the abstract specification forbids. 


Here, we investigate whether \gls{fret}'s parent-child relationship can be expressed as formal refinement. In particular, it is possible to formalise the following abstraction invariant in relation to \condition{sensorfaults}:

{\small\condition{sensorfaults} $\iff$ \condition{(sensorValue(S) $>$ nominalValue + R) $|$ (sensorValue(S) $<$ nominalValue - R) $|$ (sensorValue(S) = null)}}

\noindent Intuitively this means that the boolean \condition{sensorfaults} (from the parent requirement) corresponds to the condition on the right of the `$\iff$' (from the child requirement). This kind of refinement is referred to as \textit{data refinement}.

Similarly, the abstraction invariant between
 \condition{trackingPilotCommands} and the condition and response pair \condition{(pilotInput $=>$ setThrust = V2) \& (observedThrust = V1)} and \response{(observedThrust = V2)}
could be specified as:

\centerline{\condition{trackingPilotCommands} $\iff$ \condition{pilotInput}}

The remainder of the condition-response pair above is then treated as \textit{superposition} refinement, which adds detail during refinement.  
This approach is used because of the update of the \texttt{observedThrust} variable which is difficult to express in an abstraction invariant because it provides a behavioural update rather than a simple match between booleans. The additional \condition{when} and \timing{until} clauses in the child requirement are also superposition refinements.



The parent-child relationship in \gls{fret} appears to us to be formal refinement, at least for our set of requirements. In which case UC5\_R\_1 is refined by its three child requirements (UC5\_R\_1.1, UC5\_R\_1.2, UC5\_R\_1.3). 
We will examine this further in future work, where we will seek to translate these requirements into a formalism that supports refinement, and then examine whether the appropriate proof obligations can be discharged by theorem provers. 

\subsection{Translatable Requirements}
\label{sec: translatable}

As mentioned in \S\ref{sec:intro}, our aim is to formally verify the  aircraft engine controller system described in \S\ref{sec:controllerIntro}. It is often difficult to identify what properties a system should obey, for example what does it mean for the system to operate `correctly'. 
Identifying the properties to verify often causes difficulties for non-domain experts. FRET helped to guide conversations with the domain experts to facilitate the formalisation of the requirements.

\gls{fret} currently supports translation to the CoCoSim \cite{bourbouh2020cocosim} and Copilot \cite{perez2020copilot} verification tools. We are particularly interested in using CoCoSim since it works directly on Simulink diagrams. Thus, we have started to generate CoCoSim contracts for these requirements automatically using \gls{fret} \cite{mavridou2020bridging}. This is described in \cite{luckcuck2021verifiable} and corresponds to Phase 2A of the methodology outlined in Fig.~\ref{fig:methodology}.

As described in \S\ref{sec: patterns}, we didn't rely heavily on \timing{timing} constraints that specified specific time steps, rather we used \timing{until} constraints that could potentially be translated into boolean flags in other formalisms. As such, we believe that the vast majority of the requirements that we formalised in \gls{fret} could be used by other formal methods. For example, we may need to model the aircraft engine controller in an alternative formalism if some of these properties fail to verify using CoCoSim due to the state space explosion. This approach has been taken, manually, in previous work~\cite{bourbouh2021integrating}.

\section{Lessons Learnt and Future Improvements}
\label{sec: lessons}
This section summarises some of the lessons that we learnt from this case study. \medskip 

\noindent\textit{Communication Barrier:} We found that \gls{fret} and \fretish{} provided a useful conduit for conversation with our industrial partner. Formalising natural-language requirements is often time-consuming because of contradictions and ambiguities. \fretish{} provides a stepping-stone between readable natural-language requirements and their fully-formal counterparts, and helped us to step-wise reduce ambiguity. 
This process produced requirements that are easier to read than if they had been fully-formal, but which \gls{fret} can still automatically formalise.


We used \gls{fret} during elicitation discussions to explain and update our requirements, alongside our industrial partner.  The diagramatic semantics gave a useful visualisation of the requirements, helping both us and our industrial partner to sanity-check updates. \gls{fret} also enabled our documentation of information for each natural-language requirement, recording the reasoning for any changes, alongside each \fretish{} requirement, thus facilitating requirements explainability.\medskip

\noindent\textit{Parent-Child Relationship:} While not a formal relationship, the link between parent and child requirements enabled us to gradually make the requirements more concrete, by adding details and removing ambiguities. For example, the term \condition{sensorfaults} in UC5\_R\_1 was replaced with \condition{(sensorValue(S) $>$ nominalValue + R) $|$ (sensorValue(S) $<$ nominalValue - R) $|$ (sensorValue(S) = null)} in its child requirements (Table \ref{table: r1children}).
Documenting these links, via the `Parent ID' and `Rationale' fields in \gls{fret}, provides a structuring mechanism that enables traceability within the requirement set. However, a more concrete definition of this link would be beneficial.  We have suggested a definition using formal refinement, but an object-oriented inheritance relationship could also provide structure here. \medskip

\noindent\textit{Limitations of FRETISH:} While a useful language, we note some limitations of \fretish{}. 
Logical quantifiers ($\forall, \exists$) would be a welcome addition to \fretish{}. 
For example, in UC5\_R\_1.1, we used \condition{sensorValue(S)}, where the parameter \condition{S} indicates that this condition applies to all sensors. This is slight \textit{abuse of notation}, it would have been more accurate to use a $\forall$ quantifier.

We also suggest that highlighting assignments to variables (which hold \textit{after} the requirement is triggered) would be beneficial. For example, in UC5\_R\_1.1 we use the \texttt{observedThrust} variable in both the \Condition{} and the \Response. We expect that \texttt{observedThrust} has been updated by the \Response{} but this is not obvious, and may have implications when translating to other verification tools.\medskip

\noindent\textit{An Industrial Perspective:}
Our industrial partner had not used \gls{fret} before, so we asked them about their experience with it. They felt that the \fretish{} requirements were `\textit{much more clear}' than the natural-language requirements, and that using a `\textit{controlled-natural language with precise semantics is always better than natural-language}'. When asked  if \gls{fret} was difficult to use or understand they said that \gls{fret} was `\textit{very easy to use; interface is intuitive; formal language is adequately documented inside the tool (along with usage examples)}'. Overall, they found that \gls{fret} was useful `\textit{because it forces you to think about the actual meaning behind the natural-language requirements}'.

Having installed \gls{fret}, our industrial partner found some usability improvements that could be made. 
Some were problems with the GUI, which have a low impact but happen very frequently. 
Other issues related to \fretish{}; for example, they would like to be able to add user-defined templates and patterns, such as specifying timing within the condition component. Finally, to aid interoperability they `\textit{would like to be able to export to a format where the formalised requirements are machine readable (e.g. parse tree)}'. \medskip

\noindent\textit{Impact:} Formalising the requirements made them more detailed and less ambiguous; crucially much of the detail came from elicitation discussions, not from existing documents. \gls{fret} captures the links between requirements, and explanations of their intent (which was often more detailed than what already existed). These two things mean that the \gls{fret} requirements are a valuable development artefact. 
We are currently pursuing Phase 2 of our methodology (Fig. \ref{fig:methodology}), in which we will assess the impact of the \fretish{} requirements on verification.

We believe that \gls{fret} can scale to larger requirements sets, with the parent-child relationship providing a grouping function. However, for large sets of requirements it might be necessary to modularise or refactor the requirements so that they are easier to maintain. We are currently examining how \fretish{} requirements can be refactored for our use case.

\section{Conclusions and Future Work}
\label{sec:conclude}

This paper provides an experience report of requirements elicitation and formalisation of an aircraft engine controller in \gls{fret}. Our industrial partner provided a set of natural-language requirements, test cases, and a Simulink diagram. 
In close collaboration with our industrial partner, we clarified ambiguous text in the requirements and test cases. This was essential, as we had originally misunderstood some of the text. This iterative process produced a set of detailed \fretish{} requirements that we, and our industrial partner, are confident correspond to the intent of the natural-language requirements. 
The \fretish{} requirements are now ready for use in formal verification activities.

During this work we identified improvements that could be made to \gls{fret}, which we plan to investigate in future work. 
First, our \fretish{} requirements contain quite a lot of repetition, so if changes were needed we often had to make the change manually in several places. This was very time-consuming, so we propose adding automatic requirement refactoring.
Second, we plan to investigate how to introduce globally-declared variable types. This would improve the readability of requirements; clarifying what operations are valid for a particular variable, while encapsulating definitions that might change in the future. This could be made optional, to retain the ability to write very abstract initial requirements.
Finally, we would like to improve the interoperability of \gls{fret} with other formal verification tools. For example, adding a translator to the input language of a theorem prover to avoid the state-explosion faced by model checkers (like Kind2, which is used to verify CoCoSpec contracts); or outputting the requirement to a parse tree, as suggested by our industrial partner.

\bibliographystyle{abbrv}
\bibliography{bibliography}
\newpage

\begin{appendices}
\renewcommand{\thesection}{\appendixname~\Alph{section}}
\section{Simulink Diagram}
\label{app:simulink}

This appendix contains a high-level view of the Simulink diagram of the aircraft engine controller that we described in Sect.~\ref{sec:bg}.

\begin{figure}[h]
\centering
\rotatebox{90}{\includegraphics[scale = 0.6]{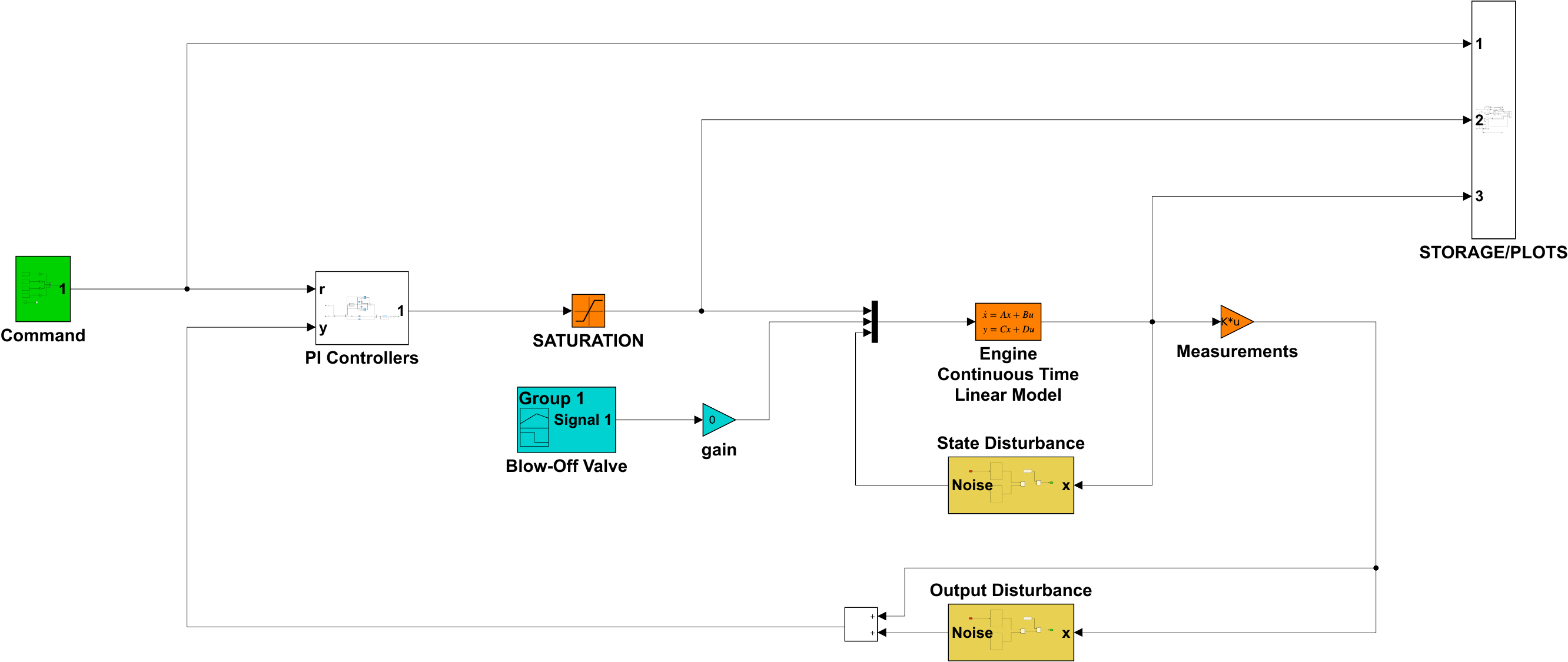}}

\caption{High-level view of the Simulink diagram that our industrial partner constructed to model the controller for this use case.}
\label{fig: simulinkl0}
\label{fig:simulinDiagram}
\end{figure}
\clearpage
\section{Test Cases}
\label{app:testCases}

This appendix contains all of the test cases for the aircraft engine controller, provided by our industry partners. There are 20 test cases in total and these are shown in Tables \ref{tab:testCases1} -- \ref{tab:testCases4}.

\input{tab-testCases.tex}
\clearpage

\section{Child Requirements}
This appendix contains the complete set of \fretish{} child requirements for the aircraft engine controller use case. These requirements are contained in Tables \ref{table: r1r2children} --\ref{table: r11r12r13r14children}.
\input{allchildreqs.tex}

\end{appendices}

\end{document}

%% file: acronyms.tex



\newacronym{ltl}{LTL}{Linear-time Temporal Logic}

\newacronym{ctl}{CTL}{Computation Tree Logic}

\newacronym{ptl}{PTL}{Probabilistic Temporal Logic}

\newacronym{pctl}{PCTL}{Probabilistic Computation Tree Logic}

\newacronym{pctl*}{PCTL*}{Probabilistic Computation Tree Logic*}

\newacronym{tctl}{TCTL}{Timed Computation Tree Logic}

\newacronym{stl}{STL}{Signal Temporal Logic}

\newacronym{ltl-x}{LTL-X}{a version of Linear Time Logic without the `next' operator}

\newacronym{mtl}{MTL}{Metric Temporal Logic}

\newacronym{iactlk-x}{IACTLK-X}{a fragment of the temporal-epistemic logic CTLK, without the `next' operator}

\newacronym{mucalc}{$\mu$-calculus}{a strict superset of other temporal logics}

\newacronym{fotl}{FOTL}{First-Order Temporal Logic}

\newacronym{bdi}{BDI}{Belief-Desire-Intention}

\newacronym{prs}{PRS}{Procedural Reasoning System}

\newacronym{are}{ARE}{Autonomous Reasoning Engine}

\newacronym{dmars}{dMARS}{distributed Multi-Agent Reasoning System}


\newacronym{fsm}{FSM}{Finite-State Machine}

\newacronym{pfsm}{PFSM}{Probabilistic Finite-State Machine}

\newacronym{apfsm}{APFSM}{\textit{Autonomous} Probabilistic Finite-State Machine}

\newacronym{pha}{PHA}{Probabilistic Hybrid Automata}

\newacronym{fsa}{FSA}{Finite-State Automata}

\newacronym{ta}{TA}{Timed Automata}

\newacronym{gspn}{GSPN}{Generalised Stochastic Petri Net}

\newacronym{mopn}{MOPN}{Marked Ordinary Petri Net}

\newacronym{ba}{BA}{B\"{u}chi Automata}

\newacronym{dtmc}{DTMC}{Discrete-Time Markov Chain}


\newacronym{pars}{PARS}{Process Algebra for Robot Schemas}

\newacronym{fsp}{FSP}{Finite State Processes}

\newacronym{piadl}{$\pi$ADL}{$\pi$ calculus combined with ADL}

\newacronym{csp}{CSP}{Communicating Sequential Processes}

\newacronym{ccs}{CCS}{Calculus of Communicating Systems}

\newacronym{wsccs}{WSCCS}{Weighted Synchronous CCS}

\newacronym{csp|b}{CSP$\|$B}{an integrated formalism combining CSP and B}


\newacronym{fdr}{FDR}{the Failures-Divergences Refinement checker}

\newacronym{jpf}{JPF}{Java PathFinder}

\newacronym{ajpf}{AJPF}{Agent Java PathFinder}

\newacronym{mcmas}{MCMAS}{Model Checker for Multi-Agent Systems}

\newacronym{ltlmop}{LTLMoP}{Linear Temporal Logic MissiOn Planning}


\newacronym{adl}{ADL}{Architecture Description Language}

\newacronym{aadl}{AADL}{Architecture Analysis and Design Language}

\newacronym{lts}{LTS}{Labelled-Transition System}

\newacronym{uml}{UML}{Unified Modelling Language}

\newacronym{sysml}{SySML}{Systems Modelling Language}

\newacronym{robotml}{RobotML}{Robot Modelling Language}

\newacronym{dsl}{DSL}{Domain Specific Language}

\newacronym{sct}{SCT}{Supervisory Control Theory}

\newacronym{ide}{IDE}{Integrated Development Environment}

\newacronym{mde}{MDE}{Model-Driven Engineering}

\newacronym{ai}{AI}{Artificial Intelligence}

\newacronym{fdir}{FDIR}{Fault Detection, Isolation and Recovery}

\newacronym{ifm}{iFM}{Integrated Formal Methods}

\newacronym{ros}{ROS}{Robot Operating System}

\newacronym{amcl}{AMCL}{Adaptive Monte Carlo Localisation}

\newacronym{dl}{\text{\upshape\textsf{d{\kern-0.05em}L}}}{differential dynamic logic}

\newacronym{vm}{VM}{Virtual Machine}

\newacronym{bip}{BIP}{Behaviour Interaction Priority}

\newacronym{fol}{FOL}{First-Order Logic}

\newacronym{owl}{OWL}{Web Ontology Language}

\newacronym{ria}{RIA}{Restore Invariant Approach}

\newacronym{ocl}{OCL}{Object Constraint Language}

\newacronym{sil}{SIL}{Safety Integrity Level}

\newacronym{rcl}{RCL}{ROS Contract Language}

\newacronym{rv}{RV}{Runtime Verification}

\newacronym{rml}{RML}{Runtime Monitoring Language}

%% file: nlreqtable.tex
\begin{table}
\begin{tabular}{|c|p{13cm}|}
\hline
\textbf{ID} & \textbf{Description}\\ \hline
UC5\_R\_1 & Under sensor faults, while tracking pilot commands, control objectives shall be satisfied (e.g., settling time, 
overshoot, and steady state error will be within predefined, acceptable limits) \\ \hline
UC5\_R\_2 & Under sensor faults, during regulation of nominal system operation (no change in pilot input), control objectives 
shall be satisfied (e.g., settling time, overshoot, and steady state error will be within predefined, acceptable limits) \\ \hline
UC5\_R\_3 & Under sensor faults, while tracking pilot commands, operating limit objectives shall be satisfied (e.g., respecting 
upper limit in shaft speed) \\ \hline
UC5\_R\_4 & Under sensor faults, during regulation of nominal system operation (no change in pilot input), operating limit 
objectives shall be satisfied (e.g., respecting upper limit in shaft speed) \\ \hline
UC5\_R\_5 & Under mechanical fatigue conditions, while tracking pilot commands, control objectives shall be satisfied (e.g., 
settling time, overshoot, and steady state error will be within predefined, acceptable limits) \\ \hline
UC5\_R\_6 & Under mechanical fatigue conditions, during regulation of nominal system operation (no change in pilot input), 
control objectives shall be satisfied (e.g., settling time, overshoot, and steady state error will be within predefined, 
acceptable limits) \\ \hline
UC5\_R\_7 & Under mechanical fatigue conditions, while tracking pilot commands, operating limit objectives shall be satisfied 
(e.g., respecting upper limit in shaft speed) \\ \hline
UC5\_R\_8 & Under mechanical fatigue conditions, during regulation of nominal system operation (no change in pilot input), 
operating limit objectives shall be satisfied (e.g., respecting upper limit in shaft speed) \\ \hline
UC5\_R\_9 & Under low probability hazardous events, while tracking pilot commands, control objectives shall be satisfied (e.g., 
settling time, overshoot, and steady state error will be within predefined, acceptable limits) \\ \hline
UC5\_R\_10 & Under low probability hazardous events, during regulation of nominal system operation (no change in pilot 
input), control objectives shall be satisfied (e.g., settling time, overshoot, and steady state error will be within 
predefined, acceptable limits) \\ \hline
UC5\_R\_11 & Under low probability hazardous events, while tracking pilot commands, operating limit objectives shall be 
satisfied (e.g., respecting upper limit in shaft speed) \\ \hline
UC5\_R\_12 & Under low probability hazardous events, during regulation of nominal system operation (no change in pilot 
input), operating limit objectives shall be satisfied (e.g., respecting upper limit in shaft speed) \\ \hline
UC5\_R\_13 & While tracking pilot commands, controller operating mode shall appropriately switch between nominal and surge 
/ stall prevention operating state \\ \hline
UC5\_R\_14 & During regulation of nominal system operation (no change in pilot input), controller operating mode shall 
appropriately switch between nominal and surge / stall prevention operating state \\ \hline
\end{tabular}
\caption{Natural-language requirements for the aircraft engine controller as produced by the aerospace use case in the VALU3S project. These 14 requirements are mainly concerned with continued operation of the controller in the presence of sensor faults (UC5\_R\_1--UC5\_R\_4), perturbation of system parameters (UC5\_R\_5--UC5\_R\_8) and low probability hazards (UC5\_R\_9--UC5\_R\_12). There are also requirements for switching between modes if engine surge/stall is detected (UC5\_R\_13--UC5\_R\_14).}
\label{table: nlreqs}
\end{table}

%% file: frethlreqs.tex
\begin{table}[t]
\begin{tabular}{|c|p{13cm}|}
\hline
\textbf{ID} & \textbf{FRETISH} \\\hline
UC5\_R\_1 & \condition{if ((sensorfaults) \& (trackingPilotCommands))} \component{Controller} shall \response{(controlObjectives)}\\\hline
UC5\_R\_2 & \condition{if ((sensorfaults) \& (!trackingPilotCommands))} \component{Controller} shall \response{(controlObjectives)} \\\hline
UC5\_R\_3 & \condition{if ((sensorfaults) \& (trackingPilotCommands))} \component{Controller} shall \response{(operatingLimitObjectives)} \\\hline
UC5\_R\_4 & \condition{if ((sensorfaults) \& (!trackingPilotCommands))} \component{Controller} shall \response{(operatingLimitObjectives)}\\\hline
UC5\_R\_5 & \condition{if ((mechanicalFatigue) \& (trackingPilotCommands))} \component{Controller} shall \response{(controlObjectives)} \\\hline
UC5\_R\_6 & \condition{if ((mechanicalFatigue) \& (!trackingPilotCommands))} \component{Controller} shall \response{(controlObjectives)}  \\\hline
UC5\_R\_7 & \condition{if ((mechanicalFatigue) \& (trackingPilotCommands))} \component{Controller} shall \response{(operatingLimitObjectives)} \\\hline
UC5\_R\_8 & \condition{if ((mechanicalFatigue) \& (!trackingPilotCommands))} \component{Controller} shall \response{(operatingLimitObjectives)}\\\hline
UC5\_R\_9 & \condition{if ((lowProbabilityHazardousEvents) \& (trackingPilotCommands))} \component{Controller} shall \response{(controlObjectives)} \\\hline
UC5\_R\_10 & \condition{if ((lowProbabilityHazardousEvents) \& (!trackingPilotCommands))} \component{Controller} shall \response{(controlObjectives)}\\\hline
UC5\_R\_11 & \condition{if ((lowProbabilityHazardousEvents) \& (trackingPilotCommands))} \component{Controller} shall \response{(operatingLimitObjectives)}\\\hline
UC5\_R\_12 & \condition{if ((lowProbabilityHazardousEvents) \& (!trackingPilotCommands))} \component{Controller} shall \response{(operatingLimitObjectives)} \\\hline
UC5\_R\_13 & \condition{if (trackingPilotCommands)} \component{Controller} shall \response{(changeMode(nominal)) $|$ (changeMode(surgeStallPrevention))}\\\hline
UC5\_R\_14 & \condition{if (!trackingPilotCommands)} \component{Controller} shall \response{(changeMode(nominal)) $|$ (changeMode(surgeStallPrevention))}\\\hline
\end{tabular}
\caption{\fretish{} parent requirements corresponding to the natural-language requirements outlined in Table \ref{table: nlreqs}. The correspondance is clear to see and we have used booleans to represent the ambiguous terms from the natural-language requirements.}
\label{table: frethlreqs}
\end{table}

%% file: r1childreqs.tex
\begin{table}[t]
\begin{tabular}{|c|c|p{11.1cm}|}
\hline
\textbf{ID} & \textbf{Parent} &\textbf{FRETISH} \\ \hline
UC5\_R\_1.1 & UC5\_R\_1 & \condition{when (diff(r(i),y(i)) $>$ E) if((sensorValue(S) $>$ nominalValue + R) $|$ (sensorValue(S) $<$ nominalValue - R) $|$ (sensorValue(S) = null) \& (pilotInput $=>$ setThrust = V2)  \& (observedThrust = V1))} \component{Controller} shall \timing{until (diff(r(i),y(i)) $<$ e)} \response{(settlingTime $>=$ 0) \& (settlingTime $<=$ settlingTimeMax) \& (observedThrust = V2)}\\\hline

UC5\_R\_1.2 & UC5\_R\_1 & \condition{when (diff(r(i),y(i)) $>$ E) if((sensorValue(S) $>$ nominalValue + R) $|$ (sensorValue(S) $<$ nominalValue - R) $|$ (sensorValue(S) = null)\& (pilotInput $=>$ setThrust = V2) \& (observedThrust = V1))} \component{Controller} shall \timing{until (diff(r(i),y(i)) $<$ e)} \response{(overshoot $>=$ 0) \& (overshoot $<=$ overshootMax) \& (observedThrust = V2)}\\\hline

UC5\_R\_1.3 & UC5\_R\_1 & \condition{when (diff(r(i),y(i)) $>$ E) if((sensorValue(S) $>$ nominalValue + R) $|$ (sensorValue(S) $<$ nominalValue - R) $|$ (sensorValue(S) = null)\& (pilotInput $=>$ setThrust = V2)\& (observedThrust = V1))} \component{Controller} shall \timing{until (diff(r(i),y(i)) $<$ e)} \response{(steadyStateError $>=$ 0) \& (steadyStateError $<=$ steadyStateErrorMax) \& (observedThrust = V2)} \\\hline
\end{tabular}
\caption{We have three distinct child requirements for UC5\_R\_1 that capture the correct behaviour with respect to each of settling time, overshoot and steady state error.}
\label{table: r1children}
\end{table}

%% file: r1314children.tex
\begin{table}[t]
\begin{tabular}{|c|c|p{10.9cm}|}
\hline
\textbf{ID} & \textbf{Parent} &\textbf{FRETISH} \\ \hline
UC5\_R\_13.1 & UC5\_R\_13 & \scope{nominal} \condition{when (diff(setNL, observedNL) $>$ NLmax) if (pilotInput $=>$ surgeStallAvoidance)} \component{Controller} shall \timing{until (diff(setNL, observedNL) $<$ NLmin)} \response{(changeMode(surgeStallPrevention)) }\\\hline

UC5\_R\_13.2 & UC5\_R\_13 & \scope{surgeStallPrevention} \condition{when (diff(setNL, observedNL) $<$ NLmax) if (pilotInput $=>$ !surgeStallAvoidance)} \component{Controller} shall \timing{until (diff(setNL, observedNL) $>$ NLmin)} \response{(changeMode(nominal)) }\\\hline

UC5\_R\_14.1 & UC5\_R\_14 & \scope{nominal} \condition{when (diff(setNL, observedNL) $>$ NLmax) if (!pilotInput $=>$ surgeStallAvoidance)} \component{Controller} shall \timing{until (diff(setNL, observedNL) $<$ NLmin)} \response{(changeMode(surgeStallPrevention)) }\\\hline

UC5\_R\_14.2 & UC5\_R\_14 & \scope{surgeStallPrevention} \condition{when (diff(setNL, observedNL) $<$ NLmax) if (!pilotInput $=>$ !surgeStallAvoidance)} \component{Controller} shall \timing{until (diff(setNL, observedNL) $>$ NLmin)} \response{(changeMode(nominal)) }\\\hline

\end{tabular}

\caption{Child requirements corresponding to UC5\_R\_13 and UC5\_R\_14. These differ from the previous requirements because we use the \Scope{} field to assert which mode of operation the controller is in.}
\label{table:fret1314children}
\end{table}

%% file: r11_1.tex
\begin{table}[t]
\begin{tabular}{|c|c|p{11cm}|}
\hline
\textbf{ID} & \textbf{Parent} &\textbf{FRETISH} \\ \hline
UC5\_R\_11.1 & UC5\_R\_11 & \condition{when (diff(r(i),y(i)) $>$ E) if (outsideAirPressure(T1) != outsideAirPressure(T2) \& (diff(t2,t1) $<$ threshold) \&(abs(outsideAirPressure (T1) - outsideAirPressure(T2)) $>$ pressureThreshold) \&(observedThrust = V1) \&(pilotInput $=>$ setThrust = V2))} \component{Controller} shall \timing{until (diff(r(i),y(i)) $<$ e)} \response{(shaftSpeed $>=$ operatingLowerBound) \& (shaftSpeed $<=$ operatingUpperBound) \& (observedThrust = V2)}\\\hline

\end{tabular}

\caption{Child requirement of UC5\_R\_11 which has timing implicit through the use of the timestamp variables \condition{T1} and \condition{T2}.}
\label{table: r11.1}
\end{table}

%% file: tab-testCases.tex
\begin{table}[!h]
\begin{tabularx}{\textwidth}{|l|l|X|}
\hline
\textbf{Test Case ID} & \textbf{Req. ID} & \textbf{Description} 
\tabularnewline
\hline

UC5\_TC\_1 & UC5\_R\_1 & \textbf{Preconditions}: Aircraft is in
operating mode M and sensor S value deviates at most +/­ R \% from
nominal value \newline
\textbf{Input conditions / steps}: Observed aircraft thrust is at value V1 and pilot input changes
from A1 to A2 \newline
\textbf{Expected results}:
Observed aircraft thrust changes and settles to value V2, respecting
control objectives (settling time, overshoot, steady state error) \tabularnewline
\hline

UC5\_TC\_2 & UC5\_R\_1 & \textbf{Preconditions}: Aircraft is in
operating mode M and sensor S value is not available (sensor is out of
order)\newline 
\textbf{Input conditions / steps}:
Observed aircraft thrust is at value V1 and pilot input changes from A1
to A2\newline 
\textbf{Expected results}: Observed
aircraft thrust changes and settles to value V2, respecting control
objectives (settling time, overshoot, steady state error) \tabularnewline \hline

UC5\_TC\_3 & UC5\_R\_2 & \textbf{Preconditions}: Aircraft is in
operating mode M and sensor S value deviates at most +/­ R \% from
nominal value \newline \textbf{Input conditions /
steps}: Observed aircraft thrust is at value V1 and perturbations in
non­pilot input cause it to change to
V2\newline \textbf{Expected results}: Observed
aircraft thrust returns to value V1, respecting control objectives
(settling time, overshoot, steady state error) \tabularnewline \hline

UC5\_TC\_4 & UC5\_R\_2 & \textbf{Preconditions}: Aircraft is in
operating mode M and sensor S value is not available (sensor is out of
order) \newline \textbf{Input conditions / steps}:
Observed aircraft thrust is at value V1 and perturbations in non­pilot
input cause it to change to
V2\newline \textbf{Expected results}: Observed
aircraft thrust returns to value V1, respecting control objectives
(settling time, overshoot, steady state error) \tabularnewline \hline

UC5\_TC\_5 & UC5\_R\_3 & \textbf{Preconditions}: Aircraft is in
operating mode M and sensor S value deviates at most +/­ R \% from
nominal value\newline \textbf{Input conditions /
steps}: Observed aircraft thrust is at value V1 and pilot input changes
from A1 to A2\newline \textbf{Expected results}:
Observed aircraft thrust changes and settles to value V2, respecting
operating limit objectives (e.g. upper limit in shaft speed) \tabularnewline \hline

UC5\_TC\_6 & UC5\_R\_3 & \textbf{Preconditions}: Aircraft is in
operating mode M and sensor S value is not available (sensor is out of
order) \newline \textbf{Input conditions / steps}:
Observed aircraft thrust is at value V1 and pilot input changes from A1
to A2\newline \textbf{Expected results}: Observed
aircraft thrust changes and settles to value V2, respecting operating
limit objectives (e.g. upper limit in shaft speed) \tabularnewline \hline

\end{tabularx}
\caption{The test cases for the aircraft engine controller, provided by our industrial partner corresponding to requirements UC5\_R\_1, UC5\_R\_2 and UC5\_R\_3. \label{tab:testCases1} }
\end{table}


\begin{table}
\begin{tabularx}{\textwidth}{|l|l|X|}
\hline
\textbf{Test Case ID} & \textbf{Req. ID} & \textbf{Description} 
\tabularnewline
\hline

UC5\_TC\_7 & UC5\_R\_4 & \textbf{Preconditions}: Aircraft is in
operating mode M and sensor S value deviates at most +/­ R \% from
nominal value\newline \textbf{Input conditions /
steps}: Observed aircraft thrust is at value V1 and perturbations in
non­pilot input cause it to change to
V2\newline \textbf{Expected results}: Observed
aircraft thrust returns to value V1, respecting operating limit
objectives (e.g. upper limit in shaft speed) \tabularnewline \hline

UC5\_TC\_8 & UC5\_R\_4 & \textbf{Preconditions}: Aircraft is in
operating mode M and sensor S value is not available (sensor is out of
order)\newline \textbf{Input conditions / steps}:
Observed aircraft thrust is at value V1 and perturbations in non­pilot
input cause it to change to
V2\newline \textbf{Expected results}: Observed
aircraft thrust returns to value V1, respecting operating limit
objectives (e.g. upper limit in shaft speed) \tabularnewline \hline

UC5\_TC\_9 & UC5\_R\_5 & \textbf{Preconditions}: Aircraft is in
operating mode M and system parameter P deviates at most +/­ R \% from
nominal value\newline \textbf{Input conditions /
steps:} Observed aircraft thrust is at value V1 and pilot input changes
from A1 to A2\newline \textbf{Expected results}:
Observed aircraft thrust changes and settles to value V2, respecting
control objectives (settling time, overshoot, steady state error) \tabularnewline \hline

UC5\_TC\_10 & UC5\_R\_6 & \textbf{Preconditions}: Aircraft is in
operating mode M and system parameter P deviates at most +/­ R \% from
nominal value\newline \textbf{Input conditions /
steps}: Observed aircraft thrust is at value V1 and perturbations in
non­pilot input cause it to change to
V2\newline \textbf{Expected results}: Observed
aircraft thrust returns to value V1, respecting control objectives
(settling time, overshoot, steady state error) \tabularnewline \hline

UC5\_TC\_11 & UC5\_R\_7 & \textbf{Preconditions}: Aircraft is in
operating mode M and system parameter P deviates at most +/­ R \% from
nominal value \newline \textbf{Input conditions /
steps}: Observed aircraft thrust is at value V1 and pilot input changes
from A1 to A2\newline \textbf{Expected results}:
Observed aircraft thrust changes and settles to value V2, respecting
operating limit objectives (e.g. upper limit in shaft speed) \tabularnewline \hline

\end{tabularx}
\caption{The test cases for the aircraft engine controller, provided by our industrial partner corresponding to requirements UC5\_R\_4, UC5\_R\_5, UC5\_R\_6 and UC5\_R\_7. \label{tab:testCases2} }
\end{table}

\begin{table}
\begin{tabularx}{\textwidth}{|l|l|X|}
\hline
\textbf{Test Case ID} & \textbf{Req. ID} & \textbf{Description} 
\tabularnewline
\hline
UC5\_TC\_12 & UC5\_R\_8 & \textbf{Preconditions}: Aircraft is in
operating mode M and system parameter P deviates at most +/­ R \% from
nominal value \newline \textbf{Input conditions /
steps}: Observed aircraft thrust is at value V1 and perturbations in
non­pilot input cause it to change to
V2\newline \textbf{Expected results}: Observed
aircraft thrust returns to value V1, respecting operating limit
objectives (e.g. upper limit in shaft speed) \tabularnewline \hline

UC5\_TC\_13 & UC5\_R\_9 & \textbf{Preconditions}: Aircraft is in
operating mode M\newline I\textbf{nput conditions /
steps}: Observed aircraft thrust is at value V1, pilot input changes
from A1 to A2, and outside air pressure abruptly changes from P1 to
P2\newline  \textbf{Expected results}: Observed
aircraft thrust changes and settles to value V2, respecting control
objectives (settling time, overshoot, steady state error) \tabularnewline \hline

UC5\_TC\_14 & UC5\_R\_10 & \textbf{Preconditions}: Aircraft is in
operating mode M\newline \textbf{Input conditions /
steps}: Observed aircraft thrust is at value V1, small perturbations in
non­pilot input cause it to change to V2, and outside air pressure
abruptly changes from P1 to P2\newline 
\textbf{Expected results}: Observed aircraft thrust returns to value V1,
respecting control objectives (settling time, overshoot, steady state
error) \tabularnewline \hline

UC5\_TC\_15 & UC5\_R\_11 & \textbf{Preconditions}: Aircraft is in
operating mode M\newline \textbf{Input conditions /
steps}: Observed aircraft thrust is at value V1, pilot input changes
from A1 to A2, and outside air pressure abruptly changes from P1 to
P2\newline \textbf{Expected results}: Observed
aircraft thrust changes and settles to value V2, respecting operating
limit objectives (e.g. upper limit in shaft speed) \tabularnewline \hline

UC5\_TC\_16 & UC5\_R\_12 & \textbf{Preconditions}: Aircraft is in
operating mode M \newline \textbf{Input conditions /
steps}: Observed aircraft thrust is at value V1, small perturbations in
non­pilot input cause it to change to V2, and outside air pressure
abruptly changes from P1 to
P2\newline \textbf{Expected results}: Observed
aircraft thrust returns to value V1, respecting operating limit
objectives (e.g. upper limit in shaft speed) \tabularnewline \hline

\end{tabularx}
\caption{The test cases for the aircraft engine controller, provided by our industrial partner corresponding to requirements UC5\_R\_8, UC5\_R\_9, UC5\_R\_10, UC5\_R\_11 and UC5\_R\_12. \label{tab:testCases3} }
\end{table}

\begin{table}
\begin{tabularx}{\textwidth}{|l|l|X|}
\hline
\textbf{Test Case ID} & \textbf{Req. ID} & \textbf{Description} 
\tabularnewline
\hline

UC5\_TC\_17 & UC5\_R\_13 & \textbf{Preconditions}: Aircraft is in
nominal operating mode\newline \textbf{Input
conditions / steps}: Pilot input changes from A1 to A2, causing surge /
stall avoidance indicator signal to be
set\newline \textbf{Expected results}: Aircraft
switches to surge / stall prevention operating mode \tabularnewline \hline

UC5\_TC\_18 & UC5\_R\_13 & \textbf{Preconditions}: Aircraft is in
surge / stall prevention operating
mode\newline I\textbf{nput conditions / steps}: Pilot
input changes from A1 to A2, causing surge / stall avoidance indicator
signal to be cleared\newline \textbf{Expected
results}: Aircraft switches to nominal operating mode \tabularnewline \hline

UC5\_TC\_19 & UC5\_R\_14 & \textbf{Preconditions}: Aircraft is in
nominal operating mode\newline \textbf{Input
conditions / steps}: Perturbations in non­pilot input cause surge /
stall avoidance indicator signal to be
set\newline \textbf{Expected results}: Aircraft
switches to surge / stall prevention operating mode \tabularnewline \hline

UC5\_TC\_20 & UC5\_R\_14 & \textbf{Preconditions}: Aircraft is in
surge / stall prevention operating
mode\newline \textbf{Input conditions / steps}:
Perturbations in non­pilot input cause surge / stall avoidance indicator
signal to be cleared\newline \textbf{Expected
results}: Aircraft switches to nominal operating mode 
\\\hline

\end{tabularx}
\caption{The test cases for the aircraft engine controller, provided by our industrial partner corresponding to requirements UC5\_R\_13, and UC5\_R\_14. \label{tab:testCases4} }
\end{table}

%% file: allchildreqs.tex
\begin{table}[h]
\begin{tabular}{|c|c|p{11cm}|}
\hline
\textbf{ID} & \textbf{Parent} &\textbf{FRETISH} \\ \hline
UC5\_R\_1.1 & UC5\_R\_1 & \condition{when (diff(r(i),y(i)) $>$ E) if((sensorValue(S) $>$ nominalValue + R) $|$ (sensorValue(S) $<$ nominalValue - R) $|$ (sensorValue(S) = null) \& (pilotInput $=>$ setThrust = V2)  \& (observedThrust = V1))} \component{Controller} shall \timing{until (diff(r(i),y(i)) $<$ e)} \response{(settlingTime $>=$ 0) \& (settlingTime $<=$ settlingTimeMax) \& (observedThrust = V2)}\\\hline

UC5\_R\_1.2 & UC5\_R\_1 & \condition{when (diff(r(i),y(i)) $>$ E) if((sensorValue(S) $>$ nominalValue + R) $|$ (sensorValue(S) $<$ nominalValue - R) $|$ (sensorValue(S) = null)\& (pilotInput $=>$ setThrust = V2) \& (observedThrust = V1))} \component{Controller} shall \timing{until (diff(r(i),y(i)) $<$ e)} \response{(overshoot $>=$ 0) \& (overshoot $<=$ overshootMax) \& (observedThrust = V2)}\\\hline

UC5\_R\_1.3 & UC5\_R\_1 & \condition{when (diff(r(i),y(i)) $>$ E) if((sensorValue(S) $>$ nominalValue + R) $|$ (sensorValue(S) $<$ nominalValue - R) $|$ (sensorValue(S) = null)\& (pilotInput $=>$ setThrust = V2)\& (observedThrust = V1))} \component{Controller} shall \timing{until (diff(r(i),y(i)) $<$ e)} \response{(steadyStateError $>=$ 0) \& (steadyStateError $<=$ steadyStateErrorMax) \& (observedThrust = V2)} \\\hline
UC5\_R\_2.1 & UC5\_R\_2 & \condition{when (diff(r(i),y(i)) $>$ E) if((sensorValue(S) $>$ nominalValue + R) $|$ (sensorValue(S) $<$ nominalValue - R) $|$ (sensorValue(S) = null)\& (!pilotInput $=>$ setThrust = V1)\& (observedThrust = V2))} \component{Controller} shall \timing{until (diff(r(i),y(i)) $<$ e)} \response{(settlingTime $>=$ 0) \& (settlingTime $<=$ settlingTimeMax) \& (observedThrust = V1)} \\\hline

UC5\_R\_2.2 & UC5\_R\_2 & \condition{when (diff(r(i),y(i)) $>$ E) if((sensorValue(S) $>$ nominalValue + R) $|$ (sensorValue(S) $<$ nominalValue - R) $|$ (sensorValue(S) = null)\& (!pilotInput $=>$ setThrust = V1)\& (observedThrust = V2))} \component{Controller} shall \timing{until (diff(r(i),y(i)) $<$ e)} \response{(overshoot $>=$ 0) \& (overshoot $<=$ overshootMax) \& (observedThrust = V1)} \\\hline 

UC5\_R\_2.3 & UC5\_R\_2 & \condition{when (diff(r(i),y(i)) $>$ E) if((sensorValue(S) $>$ nominalValue + R) $|$ (sensorValue(S) $<$ nominalValue - R) $|$ (sensorValue(S) = null)\& (!pilotInput $=>$ setThrust = V1)\& (observedThrust = V2))} \component{Controller} shall \timing{until (diff(r(i),y(i)) $<$ e)} \response{(steadyStateError  $>=$ 0) \& (steadyStateError  $<=$ steadyStateError Max) \& (observedThrust = V1)} \\\hline

\end{tabular}
\caption{Child requirements for UC5\_R\_1 and UC5\_R\_2.}
\label{table: r1r2children}
\end{table}

\begin{table}[t]
\begin{tabular}{|c|c|p{11cm}|}
\hline
\textbf{ID} & \textbf{Parent} &\textbf{FRETISH} \\ \hline
UC5\_R\_3.1 & UC5\_R\_3 & \condition{when (diff(r(i),y(i)) $>$ E) if((sensorValue(S) $>$ nominalValue + R) $|$ (sensorValue(S) $<$ nominalValue - R) $|$ (sensorValue(S) = null) \& (observedThrust = V1))\& (pilotInput $=>$ setThrust = V2)  } \component{Controller} shall \timing{until (diff(r(i),y(i)) $<$ e)} \response{(shaftSpeed $>=$ operatingLowerBound) \& (shaftSpeed $<=$ operatingUpperBound) \& (observedThrust = V2)}\\\hline

UC5\_R\_4.1 & UC5\_R\_4 & \condition{when (diff(r(i),y(i)) $>$ E) if((sensorValue(S) $>$ nominalValue + R) $|$ (sensorValue(S) $<$ nominalValue - R) $|$ (sensorValue(S) = null)\& (!pilotInput $=>$ setThrust = V1) \& (observedThrust = V2))} \component{Controller} shall \timing{until (diff(r(i),y(i)) $<$ e)} \response{(shaftSpeed $>=$ operatingLowerBound) \& (shaftSpeed $<=$ operatingUpperBound) \& (observedThrust = V1)}\\\hline

UC5\_R\_5.1 & UC5\_R\_5 & \condition{when (diff(r(i),y(i)) $>$ E) if((systemParameter(P) $>$ nominalValue + R) $|$ (systemParameter(P) $<$ nominalValue - R) $|$ (systemParameter(P) = null)\& (observedThrust = V1)\& (pilotInput $=>$ setThrust = V2))} \component{Controller} shall \timing{until (diff(r(i),y(i)) $<$ e)} \response{(settlingTime $>=$ 0) \& (settlingTime $<=$ settlingTimeMax) \& (observedThrust = V2)} \\\hline

UC5\_R\_5.2 & UC5\_R\_5 & \condition{when (diff(r(i),y(i)) $>$ E) if((systemParameter(P) $>$ nominalValue + R) $|$ (systemParameter(P) $<$ nominalValue - R) $|$ (systemParameter(P) = null)\& (observedThrust = V1)\& (pilotInput $=>$ setThrust = V2))} \component{Controller} shall \timing{until (diff(r(i),y(i)) $<$ e)} \response{(overshoot $>=$ 0) \& (overshoot $<=$ overshootMax) \& (observedThrust = V2)} \\\hline

UC5\_R\_5.3 & UC5\_R\_5 & \condition{when (diff(r(i),y(i)) $>$ E) if((systemParameter(P) $>$ nominalValue + R) $|$ (systemParameter(P) $<$ nominalValue - R) $|$ (systemParameter(P) = null)\& (observedThrust = V1)\& (pilotInput $=>$ setThrust = V2))} \component{Controller} shall \timing{until (diff(r(i),y(i)) $<$ e)} \response{(steadyStateError $>=$ 0) \& (steadyStateError $<=$ steadyStateErrorMax) \& (observedThrust = V2)} \\\hline

\end{tabular}
\caption{Child requirements for UC5\_R\_3, UC5\_R\_4 and UC5\_R\_5.}
\label{table: r3r4r5children}
\end{table}

\begin{table}[t]
\begin{tabular}{|c|c|p{11cm}|}
\hline
\textbf{ID} & \textbf{Parent} &\textbf{FRETISH} \\ \hline

UC5\_R\_6.1 & UC5\_R\_6 & \condition{when (diff(r(i),y(i)) $>$ E) if((systemParameter(P) $>$ nominalValue + R) $|$ (systemParameter(P) $<$ nominalValue - R) $|$ (systemParameter(P) = null)\& (observedThrust = V2)\& (!pilotInput $=>$ setThrust = V1))} \component{Controller} shall \timing{until (diff(r(i),y(i)) $<$ e)} \response{(settlingTime $>=$ 0) \& (settlingTime $<=$ settlingTimeMax) \& (observedThrust = V1)} \\\hline

UC5\_R\_6.2 & UC5\_R\_6 & \condition{when (diff(r(i),y(i)) $>$ E) if((systemParameter(P) $>$ nominalValue + R) $|$ (systemParameter(P) $<$ nominalValue - R) $|$ (systemParameter(P) = null)\& (observedThrust = V2)\& (!pilotInput $=>$ setThrust = V1))} \component{Controller} shall \timing{until (diff(r(i),y(i)) $<$ e)} \response{(overshoot $>=$ 0) \& (overshoot $<=$ overshootMax) \& (observedThrust = V1)} \\\hline

UC5\_R\_6.3 & UC5\_R\_6 & \condition{when (diff(r(i),y(i)) $>$ E) if((systemParameter(P) $>$ nominalValue + R) $|$ (systemParameter(P) $<$ nominalValue - R) $|$ (systemParameter(P) = null)\& (observedThrust = V2)\& (!pilotInput $=>$ setThrust = V1))} \component{Controller} shall \timing{until (diff(r(i),y(i)) $<$ e)} \response{(steadyStateError $>=$ 0) \& (steadyStateError $<=$ steadyStateErrorMax) \& (observedThrust = V1)} \\\hline

UC5\_R\_7.1 & UC5\_R\_7 & \condition{when (diff(r(i),y(i)) $>$ E) if((systemParameter(P) $>$ nominalValue + R) $|$ (systemParameter(P) $<$ nominalValue - R) $|$ (systemParameter(P) = null)\& (observedThrust = V1)\& (pilotInput $=>$ setThrust = V2))} \component{Controller} shall \timing{until (diff(r(i),y(i)) $<$ e)} \response{(shaftSpeed $>=$ operatingLowerBound) \& (shaftSpeed $<=$ operatingUpperBound) \& (observedThrust = V2)} \\\hline

UC5\_R\_8.1 & UC5\_R\_8 & \condition{when (diff(r(i),y(i)) $>$ E) if((systemParameter(P) $>$ nominalValue + R) $|$ (systemParameter(P) $<$ nominalValue - R) $|$ (systemParameter(P) = null)\& (!pilotInput $=>$ setThrust = V1)\& (observedThrust = V2))} \component{Controller} shall \timing{until (diff(r(i),y(i)) $<$ e)} \response{(shaftSpeed $>=$ operatingLowerBound) \& (shaftSpeed $<=$ operatingUpperBound) \& (observedThrust = V1)} \\\hline

\end{tabular}
\caption{Child requirements for UC5\_R\_6, UC5\_R\_7 and UC5\_R\_8.}
\label{table: r6r7r8children}
\end{table}

\begin{table}[t]
\begin{tabular}{|c|c|p{11cm}|}
\hline
\textbf{ID} & \textbf{Parent} &\textbf{FRETISH} \\ \hline

UC5\_R\_9.1 & UC5\_R\_9 & \condition{when (diff(r(i),y(i)) $>$ E) if ((outsideAirPressure(T1) != outsideAirPressure(T2) \& (diff(t2,t1) $<$ threshold) \& (abs(outsideAirPressure (T1) - outsideAirPressure(T2)) $>$ pressureThreshold) \& (observedThrust = V1)\& (pilotInput $=>$ setThrust = V2))} \component{Controller} shall \timing{until (diff(r(i),y(i)) $<$ e)} \response{(settlingTime $>=$ 0) \& (settlingTime $<=$ settlingTimeMax) \& (observedThrust = V2)} \\\hline

UC5\_R\_9.2 & UC5\_R\_9 & \condition{when (diff(r(i),y(i)) $>$ E) if ((outsideAirPressure(T1) != outsideAirPressure(T2) \& (diff(t2,t1) $<$ threshold) \& (abs(outsideAirPressure (T1) - outsideAirPressure(T2)) $>$ pressureThreshold) \& (observedThrust = V1)\& (pilotInput $=>$ setThrust = V2))} \component{Controller} shall \timing{until (diff(r(i),y(i)) $<$ e)} \response{(overshoot $>=$ 0) \& (overshoot $<=$ overshootMax) \& (observedThrust = V2)} \\\hline

UC5\_R\_9.3 & UC5\_R\_9 & \condition{when (diff(r(i),y(i)) $>$ E) if ((outsideAirPressure(T1) != outsideAirPressure(T2) \& (diff(t2,t1) $<$ threshold) \& (abs(outsideAirPressure (T1) - outsideAirPressure(T2)) $>$ pressureThreshold) \& (observedThrust = V1)\& (pilotInput $=>$ setThrust = V2))} \component{Controller} shall \timing{until (diff(r(i),y(i)) $<$ e)} \response{(steadyStateError $>=$ 0) \& (steadyStateError $<=$ steadyStateErrorMax) \& (observedThrust = V2)} \\\hline

UC5\_R\_10.1 & UC5\_R\_10 & \condition{when (diff(r(i),y(i)) $>$ E) if ((outsideAirPressure(T1) != outsideAirPressure(T2) \& (diff(t2,t1) $<$ threshold) \& (abs(outsideAirPressure (T1) - outsideAirPressure(T2)) $>$ pressureThreshold) \& (observedThrust = V2)\& (!pilotInput $=>$ setThrust = V1))} \component{Controller} shall \timing{until (diff(r(i),y(i)) $<$ e)} \response{(settlingTime $>=$ 0) \& (settlingTime $<=$ settlingTimeMax) \& (observedThrust = V1)} \\\hline

UC5\_R\_10.2 & UC5\_R\_10 &\condition{when (diff(r(i),y(i)) $>$ E) if ((outsideAirPressure(T1) != outsideAirPressure(T2) \& (diff(t2,t1) $<$ threshold) \& (abs(outsideAirPressure (T1) - outsideAirPressure(T2)) $>$ pressureThreshold) \& (observedThrust = V2)\& (!pilotInput $=>$ setThrust = V1))} \component{Controller} shall \timing{until (diff(r(i),y(i)) $<$ e)} \response{(overshoot $>=$ 0) \& (overshoot $<=$ overshootMax) \& (observedThrust = V1)} \\\hline

UC5\_R\_10.3 & UC5\_R\_10 & \condition{when (diff(r(i),y(i)) $>$ E) if ((outsideAirPressure(T1) != outsideAirPressure(T2) \& (diff(t2,t1) $<$ threshold) \& (abs(outsideAirPressure (T1) - outsideAirPressure(T2)) $>$ pressureThreshold) \& (observedThrust = V2)\& (!pilotInput $=>$ setThrust = V1))} \component{Controller} shall \timing{until (diff(r(i),y(i)) $<$ e)} \response{(steadyStateError $>=$ 0) \& (steadyStateError $<=$ steadyStateErrorMax) \& (observedThrust = V1)} \\\hline

\end{tabular}
\caption{Child requirements for UC5\_R\_9 and UC5\_R\_10.}
\label{table: r9r10children}
\end{table}

\begin{table}[t]
\begin{tabular}{|c|c|p{11cm}|}
\hline
\textbf{ID} & \textbf{Parent} &\textbf{FRETISH} \\ \hline

UC5\_R\_11.1 & UC5\_R\_11 & \condition{when (diff(r(i),y(i)) $>$ E) if ((outsideAirPressure(T1) != outsideAirPressure(T2) \& (diff(t2,t1) $<$ threshold) \& (abs(outsideAirPressure (T1) - outsideAirPressure(T2)) $>$ pressureThreshold) \& (observedThrust = V1)\& (pilotInput $=>$ setThrust = V2))} \component{Controller} shall \timing{until (diff(r(i),y(i)) $<$ e)} \response{(shaftSpeed $>=$ operatingLowerBound) \& (shaftSpeed $<=$ operatingUpperBound) \& (observedThrust = V2)} \\\hline

UC5\_R\_12.1 & UC5\_R\_12 & \condition{when (diff(r(i),y(i)) $>$ E) if ((outsideAirPressure(T1) != outsideAirPressure(T2) \& (diff(t2,t1) $<$ threshold) \& (abs(outsideAirPressure (T1) - outsideAirPressure(T2)) $>$ pressureThreshold) \& (observedThrust = V2)\& (!pilotInput $=>$ setThrust = V1))} \component{Controller} shall \timing{until (diff(r(i),y(i)) $<$ e)} \response{(shaftSpeed $>=$ operatingLowerBound) \& (shaftSpeed $<=$ operatingUpperBound) \& (observedThrust = V1)} \\\hline

UC5\_R\_13.1 & UC5\_R\_13 & \scope{nominal} \condition{when (diff(setNL, observedNL) $>$ NLmax) if (pilotInput $=>$ surgeStallAvoidance)} \component{Controller} shall \timing{until (diff(setNL, observedNL) $<$ NLmin)} \response{(changeMode(surgeStallPrevention)) }\\\hline

UC5\_R\_13.2 & UC5\_R\_13 & \scope{surgeStallPrevention} \condition{when (diff(setNL, observedNL) $<$ NLmax) if (pilotInput $=>$ !surgeStallAvoidance)} \component{Controller} shall \timing{until (diff(setNL, observedNL) $>$ NLmin)} \response{(changeMode(nominal)) }\\\hline

UC5\_R\_14.1 & UC5\_R\_14 & \scope{nominal} \condition{when (diff(setNL, observedNL) $>$ NLmax) if (!pilotInput $=>$ surgeStallAvoidance)} \component{Controller} shall \timing{until (diff(setNL, observedNL) $<$ NLmin)} \response{(changeMode(surgeStallPrevention)) }\\\hline

UC5\_R\_14.2 & UC5\_R\_14 & \scope{surgeStallPrevention} \condition{when (diff(setNL, observedNL) $<$ NLmax) if (!pilotInput $=>$ !surgeStallAvoidance)} \component{Controller} shall \timing{until (diff(setNL, observedNL) $>$ NLmin)} \response{(changeMode(nominal)) }\\\hline

\end{tabular}
\caption{Child requirements for UC5\_R\_11, UC5\_R\_12, UC5\_R\_13 and UC5\_R\_14.}
\label{table: r11r12r13r14children}
\end{table}